\documentstyle[12pt]{article}
\advance\voffset by -0.8in
\advance\hoffset by -0.55in
\newcommand{\Hil}{{\cal H}}
\newcommand{\ze}{{\bf Z}}
\newcommand{\re}{{\bf R}}
\newcommand{\K}{{\cal K}}
\newcommand{\reg}[1]{(\ref{#1})}
\begin{document}
\title{On Representations of Conformal Field
Theories and the Construction of Orbifolds}
\author{P.S. Montague \footnotemark[1]\\ Department of Applied Mathematics and
Theoretical Physics\\ University of Cambridge\\ Silver Street \\
Cambridge CB3 9EW \\ U.K.}
\maketitle
\footnotetext[1]{January-December 1995: visiting Department of Physics and
Mathematical Physics, University of Adelaide, Adelaide, SA 5005,
Australia}
\begin{abstract}
We consider representations of meromorphic bosonic chiral conformal field
theories,
and demonstrate that such a representation is completely specified by a state
within the theory. The necessary and sufficient conditions upon this state are
derived,
and, because of their form, we show that we may extend the representation to
a representation of a suitable larger conformal field theory. In particular, we
apply this
procedure to the lattice (FKS) conformal field theories, and deduce that Dong's
proof of the uniqueness of the twisted representation for the
reflection-twisted
projection of the Leech lattice conformal field theory generalises to an
arbitrary
even (self-dual) lattice. As a consequence, we see that the reflection-twisted
lattice
theories of \cite{DGMtriality} are truly self-dual, extending the analogies
with the theories
of lattices and codes which were being pursued.
Some comments are also made on the general concept
of the definition of an orbifold of a conformal field theory in relation to
this point
of view.
\end{abstract}
\vfill
\eject
\section{Introduction}
In this letter, we shall study only meromorphic bosonic conformal field
theories.
The classification, in particular, of the self-dual theories has been a major
goal
in recent years, partly because of its physical relevance to heterotic string
theory
\cite{Schell:Venkov}, but also because of its intrinsic interest, particularly
in view
of the surprising links to the Monster group \cite{FLMbook}. In
\cite{DGMtriality,DGMtrialsumm}
these links were extended as part of a program of studying the apparent and, as
it
turned out, deep connections and analogies between bosonic conformal field
theories
and the theories of lattices and codes. One aspect of the work of
\cite{DGMtriality} and
similarly \cite{FHL} is the attempt to systematise and axiomatise the study of
these theories.
In the spirit of this approach, we study here in section \ref{defs}
the representations of such conformal field
theories, showing that they may be described entirely within the
original theory in terms of a certain corresponding state,
and as a consequence of this point of view we present, in section
\ref{apps},
a technique by
which one may extend a representation of a conformal field theory to a
theory in which it is suitably embedded and hence we tighten up the analogies
between lattices and conformal
field theories set out in \cite{DGMtrialsumm} by demonstrating
that the reflection-twisted orbifolds
of the FKS lattice conformal field theories are self-dual in the sense of being
their own
unique meromorphic representation rather than simply requiring that their
partition
function be modular invariant. Finally, in section \ref{concs}, we
present some comments and conclusions.
\section{Representations of Meromorphic Bosonic Conformal Field
  Theories}
\label{defs}
We take $\Hil$ to be a meromorphic bosonic conformal field theory with vertex
operators
denoted by $V(\psi,z)$, $\psi\in\Hil$, and consider
a representation of $\Hil$ given by a set of vertex operators $U$ and a Hilbert
space $\K$ (see \cite{DGMtriality}
for a precise definition of these concepts), {\em i.e.}
\begin{equation}
U(\psi,z)U(\phi,w)=U(V(\psi,z-w)\phi,w)\,,
\label{repdef}
\end{equation}
for $\psi$, $\phi\in\Hil$, and also $U(|0\rangle,z)\equiv 1$ (where $|0\rangle$
denotes
the vacuum state in $\Hil$ -- the unique state invariant under the SU(1,1)
symmetry
generated by the Virasoro modes $L_0$, $L_{\pm 1}$).

Now, the existence of the representation is equivalent to the existence of
intertwining
operators \cite{DGMtriality} $W(\chi,z)$, $\chi\in K$, defined by
$W(\chi,z)\psi=
e^{zL_{-1}}U(\psi,-z)\chi$, such that we have the ``intertwining" locality
relation
\begin{equation}
U(\psi,z)W(\chi,w)=W(\chi,w)V(\psi,z)\,,
\end{equation}
where the left and right hand sides of the relation are strictly only defined
for $|z|>|w|$ and $|w|>|z|$ respectively, and the equality is taken to mean
that
when we take matrix elements of both sides in the appropriate Hilbert spaces
the
functions that are so obtained on either side may be analytically continued
into one
another.

We shall also require that our conformal field theory $\Hil$ possess a
hermitian
structure \cite{DGMtriality}, by which we mean that there exists an
antilinear
map on $\psi\mapsto\overline\psi$ on $\Hil$ such that
\begin{equation}
V(\overline\psi,z)=V\left(e^{z^\ast L_1}{z^\ast}^{-2L_0}\psi,1/z^\ast
\right)^\dagger\,,
\end{equation}
and we shall say that the representation is hermitian if a
similar relation holds for the vertex operators $U$ for some
antilinear map on $\K$.

Consider the matrix element
\begin{equation}
\langle\chi|U(\psi_1,z_1) U(\psi_2,z_2) \ldots
U(\psi_n,z_n)|\chi\rangle
\end{equation}
for some quasi-primary ({\em i.e.} annihilated by $L_1$)
state $\chi\in\K$ and states $\psi_1,\ldots\psi_n\in\Hil$.
Using the definition of the intertwining operators, this becomes
\begin{equation}
\langle\chi|U(\psi_1,z_1) U(\psi_2,z_2) \ldots
U(\psi_{n-1},z_{n-1})e^{z_nL_{-1}}W(\chi,-z_n)|\psi_n\rangle\,,
\end{equation}
and then using the fact that $L_{-1}$ is the generator of translations
we obtain
\begin{equation}
\langle\chi|U(\psi_1,z_1-z_n) U(\psi_2,z_2-z_n) \ldots
U(\psi_{n-1},z_{n-1}-z_n)W(\chi,-z_n)|\psi_n\rangle\,,
\end{equation}
and thus the intertwining locality relation gives
\begin{eqnarray}
\langle\chi|U(\psi_1,z_1) U(\psi_2,z_2)&\ldots&
U(\psi_n,z_n)|\chi\rangle=\nonumber\\
\langle P(z_n^\ast)|V(\psi_1,z_1-z_n) V(\psi_2,z_2-z_n)&\ldots&
V(\psi_{n-1},z_{n-1}-z_n)|\psi_n\rangle\,,
\label{maindef}
\end{eqnarray}
where
\begin{equation}
\langle P(z^\ast)|\equiv \langle\chi|W(\chi,-z)\,.
\label{Pdef}
\end{equation}
Now, in view of this, let us start again, and take \reg{maindef} as
defining an abstract representation described by vertex operators
$U$ of the conformal field theory $\Hil$. We shall investigate what
properties the now unknown state $P(z)\equiv\sum_{n\geq 0}P_nz^{-n}$
($P_n$ of conformal weight $n$)
must satisfy in order that this
be a representation. This representation is built up on the state
$\chi$ by the action of the $U$'s.
Note that, in order to make sense of this in Hilbert space terms, we
shall require that the representation be hermitian, so that the
interpretation in terms of matrix elements is simplified.
There is a vast degeneracy in this notation, and in order to identify
the Hilbert space behind the representation one must quotient out by
all the null states (identified using \reg{maindef}). Any technical
problems to do with separability of the resulting space will not be
addressed here, though in any given case it will be easy to identify
that there is actually a {\em finite} number of states at each
conformal weight.

Let us first consider the representation property \reg{repdef} and
impose that \reg{maindef} satisfy this. There are two classes of cases to
consider -- ones involving $\psi_n$ and ones not. For the first case:
\begin{eqnarray}
&&\langle\chi|U(\psi_1,z_1)\ldots U(\psi_i,z_i)U(\psi_{i+1},z_{i+1})\ldots
U(\psi_n,z_n)|\chi\rangle\nonumber\\
&=&\langle P(z_n^\ast)|V(\psi_1,z_1-z_n)\ldots
V(\psi_i,z_i-z_n)V(\psi_{i+1},z_{i+1}-z_n)\ldots
V(\psi_{n-1},z_{n-1}-z_n)|\psi_n\rangle\nonumber\\
&=&\langle P(z_n^\ast)|V(\psi_1,z_1-z_n)\ldots
V\left(V(\psi_i,z_i-z_{i+1})\psi_{i+1},
z_{i+1}-z_n\right)\ldots
V(\psi_{n-1},z_{n-1}-z_n)|\psi_n\rangle\nonumber\\
&=&\langle\chi|U(\psi_1,z_1)\ldots U\left(V(\psi_i,z_i-z_{i+1})\psi_{i+1},
z_{i+1}\right)\ldots U(\psi_n,z_n)|\chi\rangle\,,
\end{eqnarray}
and for the second case:
\begin{eqnarray}
&&\langle\chi|U(\psi_1,z_1)\ldots U(\psi_n,z_n)|\chi\rangle\nonumber\\
&=&\langle P(z_n^\ast)|V(\psi_1,z_1-z_n)\ldots
V(\psi_{n-1},z_{n-1}-z_n)|\psi_n\rangle\nonumber\\
&=&\langle\chi|U(\psi_1,z_1)\ldots
U\left(V(\psi_{n-1},z_{n-1}-z_n)\psi_n,z_n\right)|\chi\rangle\,.
\end{eqnarray}
Now, $U(|0\rangle,z)=1$ is clearly consistent for $\psi_i=|0\rangle$ with
$i\neq n$.
For $i=n$, there are two cases -- $n=1$ and $n>1$.

\noindent
$n=1$:
\begin{equation}
\langle P(z^\ast)|0\rangle=1\,.
\end{equation}
$n>1$:
\begin{eqnarray}
&&\langle P(z_n^\ast)|V(\psi_1,z_1-z_n)\ldots
V(\psi_{n-1},z_{n-1}-z_n)|0\rangle\nonumber\\
&=&\langle P(z_n^\ast)|V(\psi_1,z_1-z_n)\ldots
e^{(z_{n-1}-z_n)L_{-1}}|\psi_{n-1}\rangle\nonumber\\
&=&\langle
%% FOLLOWING LINE CANNOT BE BROKEN BEFORE 80 CHAR
P(z_n^\ast)|e^{(z_{n-1}-z_n)L_{-1}}V(\psi_1,z_1-z_{n-1})\ldots|\psi_{n-1}\rangle\,.
\end{eqnarray}
Hence, we require
\begin{equation}
\langle P(z^\ast)|=\langle P(w^\ast)|e^{(z-w)L_{-1}}\,.
\label{L1rel}
\end{equation}
[Note that this is simply the usual $L_{-1}$ commutation relation for
vertex operators when we rewrite
it in terms of $W$ using \reg{Pdef}.]

Now let us consider the hermitian structure of the representation.
Consider
\begin{eqnarray}
&&\langle\chi|U\left(e^{z_n^\ast
L_1}{z_n^\ast}^{-2L_0}\overline{\psi_n},1/z_n^\ast\right)
\ldots|\chi\rangle^\ast\nonumber\\
%% FOLLOWING LINE CANNOT BE BROKEN BEFORE 80 CHAR
&=&\langle\overline{\psi_1}|{z_1}^{-2L_0}e^{z_1L_{-1}}V^\dagger\left(e^{z_2^\ast L_1}
{z_2^\ast}^{-2L_0}\overline{\psi_2},{1\over z_2^\ast}-{1\over
z_1^\ast}\right)\dots|P(1/z_1)\rangle\nonumber\\
&=&\langle\overline{\psi_1}|{z_1}^{-2L_0}e^{z_1L_{-1}}V\left(
e^{\left({1\over z_2}-{1\over z_1}\right)L_1}\left({1\over z_2}-{1\over
z_1}\right)^{-2L_0}
e^{-z_2L_1}{z_2}^{-2L_0}\psi_2,{z_1z_2\over z_1-z_2}\right)\ldots
|P(1/z_1)\rangle\nonumber\\
&=&\langle\overline{\psi_1}|V\left(e^{(z_1-z_2)L_1}(z_1-z_2)^{-2L_0}\psi_2,
{1\over z_1-z_2}\right)\ldots{z_1}^{-2L_0}e^{z_1L_{-1}}
|P(1/z_1)\rangle\nonumber\\
&=&\langle P(1/z_1)|e^{z_1^\ast L_1}{z_1^\ast}^{-2L_0}
V(\overline{\psi_n},z_1^\ast-z_n^\ast)\ldots|\overline{\psi_1}\rangle^\ast\,.
\end{eqnarray}
We require this, for hermiticity, to be equal to
\begin{equation}
\langle P(z_1^\ast)|V(\psi_n, z_n-z_1)\ldots|\psi_1\rangle\,,
\end{equation}
using locality in the representation to reorder the operators as appropriate.
This is, using the {\em reality} of the underlying conformal field
theory (see below),
\begin{equation}
%% FOLLOWING LINE CANNOT BE BROKEN BEFORE 80 CHAR
\langle\overline{P(z_1^\ast)}|V(\overline{\psi_n},z_1^\ast-z_n^\ast)\ldots|\overline{
\psi_1}\rangle^\ast\,.
\end{equation}
Hence, we have a hermitian structure in the representation if
\begin{equation}
\langle\overline{P(z)}|=\langle P(1/z^\ast)|e^{zL_1}{z}^{-2L_0}\,.
\label{hermcond}
\end{equation}
A further property which we may wish to impose on the representation
is that of reality. The representation is said to be real if
there is an antilinear map
$\rho\mapsto\overline\rho$ on $\K$ such that
$\overline{\overline\rho}=\rho$, $L_{-1}\overline\rho=-\overline{L_{-1}\rho}$
and
\begin{equation}
(f_{\rho_1\phi\rho_2})^\ast=(-1)^{h_1+h_\phi+h_2}f_{\overline\rho_1
\overline\phi
\overline\rho_2}\,,
\label{freln}
\end{equation}
where $L_0\rho_j=h_j\rho_j$ for $j=1$, $2$,
$L_0\phi=h_\phi\phi$ and
$f_{\rho_1\phi\rho_2}=\langle\overline\rho_1|U(\phi,1)|\rho_2\rangle$. Also,
we require that if $L_0\rho=h_\rho\rho$ then
$L_0\overline\rho=h_\rho\overline\rho$. An analogous result holds true
for $\Hil$ \cite{DGMtriality}
simply as a consequence of its hermitian structure (as we
made use of above). Reality for the representation is a necessary
requirement if we wish to endow $\Hil\oplus\K$ with the structure of a
conformal field theory \cite{DGMtriality}, {\em i.e.} it is relevant
in the construction of $\ze_2$-orbifold theories.

Using the required structure as a paradigm,
we define
\begin{equation}
\overline{U(\psi,z)\chi}=U(\overline\psi,-z^\ast)\chi\,.
\end{equation}
(Note that we are implicitly taking the state $\chi$ to be real.)
This trivially satisfies the requirements of antilinearity,
$\overline{\overline\rho}=\rho$, $L_{-1}\overline\rho=-\overline{L_{-1}\rho}$
and $L_0\overline\rho=h_\rho\overline\rho$, since (denoting the state
corresponding to the Virasoro modes as $\psi_L(=\overline{\psi_L})$, {\em i.e.}
$V(\psi_L,z)\equiv\sum_nL_nz^{-n-2}$)
\begin{eqnarray}
\overline{U(\psi_L,z)U(\psi,w)\chi}&=&\overline
{U(V(\psi_L,z-w)\psi,w)\chi}\nonumber\\
&=&U(\overline{V(\psi_L,z-w)\psi},-w^\ast)\chi\nonumber\\
&=&U(V(\psi_L,w^\ast-z^\ast)\overline\psi,-w^\ast)\chi\nonumber\\
&=&U(\psi_L,-z^\ast)U(\overline\psi,-w^\ast)\chi\nonumber\\
&=&U(\psi_L,-z^\ast)\overline{U(\psi,w)\chi}\,.
\end{eqnarray}
Finally, in order to satisfy \reg{freln}, we require
\begin{equation}
\langle\rho_2|U^\dagger(\phi,1)|\overline{\rho_1}\rangle=
\langle\rho_1|U(\overline\phi,-1)|\overline{\rho_2}\rangle\,,
\end{equation}
or
\begin{equation}
\langle\rho_2|U\left(e^{L_1}\overline\phi,1\right)|\overline{\rho_1}\rangle=
\langle\rho_1|U(\overline\phi,-1)|\overline{\rho_2}\rangle\,,
\end{equation}
{\em i.e.}
\begin{equation}
\langle\rho|U^\dagger(\psi_2,z_2)U\left(e^{L_1}\overline\phi,1\right)\overline{
U(\psi_1,z_1)|\chi\rangle}=
\langle\chi|U^\dagger(\psi_1,z_1)U(\overline\phi,-1)\overline{
U(\psi_2,z_2)|\chi\rangle}\,.
\end{equation}
Thus, we require
\begin{eqnarray}
&&\langle\chi|U\left(e^{z_2^\ast
  L_1}{z_2^\ast}^{-2L_0}\overline{\psi_2},1/z_2^\ast\right)
%% FOLLOWING LINE CANNOT BE BROKEN BEFORE 80 CHAR
U\left(e^{L_1}\overline\phi,1\right)U(\overline{\psi_1},-z_1^\ast|\chi\rangle\nonumber\\
&=&\langle\chi|U\left(e^{z_1^\ast
L_1}{z_1^\ast}^{-2L_0}\overline{\psi_1},1/z_1^\ast\right)
U(\overline\phi,-1)U(\overline{\psi_2},-z_2^\ast)|\chi\,.
\end{eqnarray}
The right hand side is, by definition,
\begin{equation}
\langle P(-z_2)|V\left(e^{z_1^\ast
L_1}{z_1^\ast}^{-2L_0}\overline{\psi_1},1/z_1^\ast+z_2^\ast
\right)V(\overline\phi,-1+z_2^\ast)|\overline{\psi_2}\rangle\,,
\end{equation}
while the left hand side is
\begin{eqnarray}
%% FOLLOWING LINE CANNOT BE BROKEN BEFORE 80 CHAR
&&\langle\chi|U\left(e^{-z_1L_1}{z_1}^{-2L_0}\psi_1,-1/z_1\right)U(\phi,1)U(\psi_2,z_2)|\chi\rangle
\nonumber\\
&=&\langle
%% FOLLOWING LINE CANNOT BE BROKEN BEFORE 80 CHAR
P(z_2^\ast)|V\left(e^{-z_1L_1}{z_1}^{-2L_0}\psi_1,-1/z_1-z_2\right)V(\phi,1-z_2)|\psi_2
\rangle\,.
\end{eqnarray}
So, we see again from reality of the underlying conformal field theory that we
have a
real representation if
\begin{equation}
\langle\overline{P(z)}|=\langle P(-z^\ast)|\,.
\end{equation}

Collecting together the above results, we see that \reg{maindef}
defines a hermitian representation of $\Hil$ if
\begin{eqnarray}
\langle P(z^\ast)|0\rangle&=&1\label{first}\\
\langle P(z^\ast)|&=&\langle P(w^\ast)|e^{(z-w)L_{-1}}\label{second}\\
\langle\overline{P(z)}|&=&\langle P(1/z^\ast)|e^{zL_1}{z}^{-2L_0}
\,.\label{third}
\end{eqnarray}
If, further, we require this representation to be real then we may
replace
\reg{third} by
\begin{eqnarray}
\langle\overline{P(z)}|&=&\langle P(-z^\ast)|\label{onethird}\\
\langle P(-z^\ast)|&=&\langle
P(1/z^\ast)|e^{zL_1}{z}^{-2L_0}\,,\label{twothird}
\end{eqnarray}
where \reg{twothird} is the usual skew-symmetry relation on the
vertex operators \cite{FHL}.

Thus, any given representation of a conformal field theory is
completely specified by a state within that theory (knowledge of
$\langle P(1)|$ is clearly sufficient), and no explicit form for the
vertex operators acting in the representation space need be considered.
\section{Extension of a Representation:
An Application to FKS Lattice Conformal Field Theories}
\label{apps}
Let $\Lambda$ be an even self-dual lattice of dimension $d$,
and $\Hil(\Lambda)$ the corresponding
FKS conformal field theory (physically this represents the propagation
of a bosonic string on the torus $\re^d/\Lambda$) - constructed by the action
of a set of bosonic creation and annihilation operators, $a^i_n$, $1\leq i
\leq d$, $n\in\ze$, satisfying canonical commutation relations, on momentum
states $|\lambda\rangle$, $\lambda\in\Lambda$. Such theories admit
an involution given by lifting the reflection symmetry of the
lattice to the conformal field theory.
Let us write $\Hil(\Lambda)=\Hil(\Lambda)_+\oplus
\Hil(\Lambda)_-$, where $\Hil(\Lambda)_+$ is the sub-theory invariant
under this automorphism.
In \cite{DGMtwisted}, a space $\Hil_T(\Lambda)$ (and a set of vertex operators
acting
upon it) is constructed by acting with half-integrally graded bosonic
creation and annihilation operators upon a ground state forming a
representation for a gamma matrix algebra related to the lattice.
This forms (when $d$ is a multiple of 8)
a non-meromorphic representation of $\Hil(\Lambda)$ to
which the involution on $\Hil(\Lambda)$ lifts. Projecting out by this
involution, we obtain a meromorphic representation $\Hil_T(\Lambda)_+$
of $\Hil(\Lambda)_+$. It is further demonstrated in \cite{DGMtwisted}
that
$\widetilde\Hil(\Lambda)\equiv\Hil(\Lambda)_+\oplus\Hil_T(\Lambda)_+$
may be given a consistent structure as a conformal field theory
(provided that $\sqrt 2\Lambda^\ast$ is an even lattice).
Now, a conformal field theory is said to be self-dual if the partition
function is modular invariant (at least up to a phase under
$T:\tau\mapsto\tau+1$), and it can be shown that $\Hil(\Lambda)$ and
$\widetilde\Hil(\Lambda)$ are self-dual if and only if $\Lambda$ is
self-dual.
These are so far the only self-dual theories which have been
explicitly constructed.

In view of the analogies begun in \cite{PGmer} and developed in
\cite{DGMtrialsumm} between codes, lattices and conformal field
theories, this definition of the self-duality of a conformal field
theory is rather unsatisfactory. We would prefer to say that a theory
is self-dual if and only if it forms its own unique meromorphic
representation \cite{PGmer}. We now show that these two definitions
are equivalent in the case of the theories
$\widetilde\Hil(\Lambda)$ (the result for $\Hil(\Lambda)$ has already
been established in \cite{thesis}).

Suppose that we are given a meromorphic representation $U$ of
$\Hil(\Lambda)_+$. We construct a corresponding state $\langle
P_U(z)|$ in $\Hil(\Lambda)_+$ as at the beginning of section
\ref{defs}. The properties satisfied by this state as a consequence of
it defining a representation are such that it also defines a
representation of $\Hil(\Lambda)$. In general,
(\ref{first}-\ref{third})
(or (\ref{first}-\ref{second}) and (\ref{onethird}-\ref{twothird}))
are such that any representation of a sub-theory of a conformal field
theory may be extended in this way to a representation of the larger
theory provided the Virasoro operators $L_0$ and $L_{\pm 1}$ are
identical in both theories. In particular, we may extend any
representation of a {\em sub-conformal field theory}
\cite{DGMtriality} to one of the full theory.

However, there is no need for the representation of $\Hil(\Lambda)$
which we have so produced to be meromorphic. All that we require is
that we may restrict it to a meromorphic representation of
$\Hil(\Lambda)_+$ (specifically $U$). The twisted non-meromorphic
representations of $\Hil(\Lambda)$ ({\em i.e.} those non-meromorphic
representations such that $U\left(a_{-1}^i|0\rangle,e^{2\pi i}z\right)
=\omega U(a^i_{-1}|0\rangle,z)$ for some constant phase $\omega$ for
the weight one states $a_{-1}^i|0\rangle$ corresponding to the Cartan
subalgebra) are easily classified
\cite{dongtwisrep}. Note however that these may not correspond to the full
set of non-meromorphic representations, though they are all that we require
here.
They are simply either cosets of $\Hil(\Lambda)$ in $\Hil(\Lambda^\ast)$
or
are built up by the action of a set of
bosonic creation and annihilation operators graded by $\ze+r$ ($0<r<1$)
on a
ground state representing a gamma matrix algebra related to
$\Lambda$.
We require that the representation restrict to a meromorphic representation
of $\Hil(\Lambda)_+$.
It is clear from the fact that $L_{-1}$ generates translations, i.e.
that $[L_{-1},U(\psi,z)]={d\over dz}U(\psi,z)$,
that any non-meromorphic representation of $\Hil(\Lambda)$ must be such that
$U\left(a_{-1}^i|0\rangle,e^{2\pi i}z\right)
=\omega^iU(a^i_{-1}|0\rangle,z)$, for some phases $\omega^i$. Then the
operator product expansion \cite{PGmer}
\begin{equation}
U(a^i_{-1}|0\rangle,z)U(a^j_{-1}|0\rangle,w)=\delta^{ij}(z-w)^{-2}
+U(a_{-1}^ia_{-1}^j|0\rangle,w)+O(z-w)\
\end{equation}
implies that either $\omega^i=1$ for all $i$ or
$\omega^i=-1$ for all $i$, since $U(a_{-1}^ia_{-1}^j|0\rangle,w)$ is required
to be meromorphic.
Hence we must
have either $\Hil(\Sigma)$, for cosets $\Sigma$ of $\Lambda$ in
$\Lambda^\ast$, or
$\Hil_T(\Lambda)$ (with inequivalent representations being determined
by inequivalent representations of the relevant gamma matrix algebra - there
being a unique such representation only for $\Lambda$ self-dual),
and so $U$ is either
$\Hil(\Sigma)_\pm$ or $\Hil_T(\Lambda)_+$.

If we consider an irreducible
representation of $\widetilde\Hil(\Lambda)$, then in
particular it must decompose into irreducible representations of
$\Hil(\Lambda)_+$. Considering the action of $\Hil_T(\Lambda)_+$, we
then see from the list of possibilities above that the only solution
when $\Lambda$ is self-dual is $\Hil(\Lambda)_+\oplus\Hil_T(\Lambda)_+$
(and that there are more solutions when $\Lambda$ is not self-dual).
Thus we have shown that $\widetilde\Hil(\Lambda)$ is truly self-dual
if and only if $\Lambda$ is self-dual.
\section{Comments and Conclusions}
There is clearly much more to be done with regard to this perspective
on the representation theory of bosonic meromorphic conformal field theories.
In particular, we would wish to classify in some way all solutions of the
equations (\ref{first}-\ref{third}) (and thus give a classification of the
representations)
for some class of conformal field theories. However, the restrictions on the
state
$\chi$ ({\em i.e.} that it be real and quasi-primary) are not sufficient to
specify
$\langle P(z^\ast)|$ uniquely, and we either have to identify some sort of
cohomological
equivalence classes into which the distinct states group or restrict $\chi$
further,
{\em e.g.} by imposing that it be a ground state of the representation
(annihilated
by all positive modes of the vertex operators in the
representation space). This is the subject of ongoing research. While this
paper was in preparation, work of Zhu \cite{Zhu} came to the attention
of the author in which related techniques are employed. Zhu demonstrates
that the meromorphic representations of a vertex operator algebra $V$
are in one-to-one correspondence with representations of a certain
algebra $A(V)$ which he constructs, though his approach is difficult to
employ in practise since the algebra $A(V)$ is rather tricky to
construct, and no consideration is given to
non-meromorphic representations. However,
Zhu effectively assumes that the state
$\chi$ in our notation is a highest weight state in the representation,
and shows that this means that it is annihilated by the action of the
zero modes of the vertex operators corresponding to states in
what he denotes by $O(V)$. This should allow us to restrict our state
$\langle P(1)|$ further ($\langle P(1)|$ must be orthogonal to
$O(V)$) and the consequences of this in specific examples form
the subject of current research.

In addition, a general proof of the equivalence of the distinct definitions of
self-duality
needs to be obtained, rather than having to establish this result separately
for each class of theories considered.

We note that the technique of extending a meromorphic representation to a
non-meromorphic representation of a larger conformal field theory helps us
make more sense of a general abstract definition of an orbifold, as opposed
to the usual geometric interpretation in terms of strings propagating on
singular manifolds. Given an automorphism $\theta$ of a conformal field theory
and a representation of the $\theta=1$ subspace, we may extend this
representation
to one of the full conformal field theory (the Virasoro modes are
invariant under the action of the automorphism \cite{thesis}, and so
the conditions required for the extension by the method of the last
section to exist are therefore met). We require $U(\psi,e^{2\pi i}z)=
e^{2\pi ir\over n}U(\psi,z)$ for $\theta\psi=e^{2\pi ir\over n}\psi$ for
consistency with
the definition of an orbifold. Thus, we are able to say how a particular
meromorphic
representation corresponds to a certain non-meromorphic larger structure.
Note though that we still have no means of showing that the orbifold is unique,
beyond
actually demonstrating uniqueness of solutions to the representation equations
(\ref{first}-\ref{third}) in a certain equivalence class.
It should be noted that to actually verify that the orbifold is consistent is
still an intricate
problem which is not made any simpler in this approach. For example, in the
case of
$\widetilde\Hil(\Lambda)$, we can easily show that a $\langle P(z^\ast)|$
corresponding
to $\Hil_T(\Lambda)_+$ gives us a consistent representation of
$\Hil(\Lambda)_+$,
but to verify the final locality relation for consistency of
$\widetilde\Hil(\Lambda)$ \cite{DGMtwisted} requires a detailed check
on matrix elements of the form
\begin{equation}
\langle P(z^\ast)|V(\psi_1,z_1)\ldots V(\psi_n,z_n)|P(w^\ast)\rangle\,.
\end{equation}
Since we find that $\sqrt2\Lambda^\ast$ must be even in order that this
locality relation holds
\cite{DGMtwisted}, then it is not easy to see how there can be any
simple requirement on $\langle
P(z^\ast)|$.
Nevertheless,  we feel that this point of view provides a firm foundation on
which
to consider both orbifolds and representations of conformal field
theories from a more abstract perspective, less constrained by any
requirement for an explicit construction.
\label{concs}
\section{Acknowledgements}
The author is grateful to Gonville and Caius College, Cambridge, for a
Research Fellowship and to the Royal Society for a Commonwealth
Fellowship at the University of Adelaide. He would also like to thank
Peter Goddard for useful conversations and encouragement.
\vfill
\eject

\end{document}